\documentclass[aps,prl,preprint,tightenlines,superscriptaddress,showpacs,byrevtex]{revtex4}
\usepackage{graphicx} 
\usepackage{dcolumn}  
\graphicspath{{ps}}

\usepackage{color}
\newcommand{\notM}{\mu\!\!\!/}
\newcommand{\notE}{e\!\!\!/}
\newcommand{\TMG}{\tau\rightarrow\mu\gamma}
\newcommand{\TEG}{\tau\rightarrow{e}\gamma}
\newcommand{\MG}{\mu\gamma}
\newcommand{\EG}{{e}\gamma}
\newcommand{\fbi}{fb${}^{-1}$}

\newcommand{\gev}{{GeV}}

\newcommand{\gevpcs}{{GeV$/c^2$}}
\newcommand{\mev}{{MeV}}

\newcommand{\mevpcs}{{MeV$/c^2$}}

\newcommand{\notmu}{\mu\!\!\!/}
\newcommand{\minv}{M_{\rm inv}}
\newcommand{\dE}{{\mit \Delta} E}

\begin{document}

\preprint{\vbox{ \hbox{   }
                 \hbox{BELLE-CONF-0653}
}}

\title{ \quad\\[0.5cm]
A New Search for $\tau \to \mu \gamma $ and $\tau \to e \gamma$ Decays at Belle
}

\affiliation{Budker Institute of Nuclear Physics, Novosibirsk}
\affiliation{Chiba University, Chiba}
\affiliation{Chonnam National University, Kwangju}
\affiliation{University of Cincinnati, Cincinnati, Ohio 45221}
\affiliation{University of Frankfurt, Frankfurt}
\affiliation{The Graduate University for Advanced Studies, Hayama} 
\affiliation{Gyeongsang National University, Chinju}
\affiliation{University of Hawaii, Honolulu, Hawaii 96822}
\affiliation{High Energy Accelerator Research Organization (KEK), Tsukuba}
\affiliation{Hiroshima Institute of Technology, Hiroshima}
\affiliation{University of Illinois at Urbana-Champaign, Urbana, Illinois 61801}
\affiliation{Institute of High Energy Physics, Chinese Academy of Sciences, Beijing}
\affiliation{Institute of High Energy Physics, Vienna}
\affiliation{Institute of High Energy Physics, Protvino}
\affiliation{Institute for Theoretical and Experimental Physics, Moscow}
\affiliation{J. Stefan Institute, Ljubljana}
\affiliation{Kanagawa University, Yokohama}
\affiliation{Korea University, Seoul}
\affiliation{Kyoto University, Kyoto}
\affiliation{Kyungpook National University, Taegu}
\affiliation{Swiss Federal Institute of Technology of Lausanne, EPFL, Lausanne}
\affiliation{University of Ljubljana, Ljubljana}
\affiliation{University of Maribor, Maribor}
\affiliation{University of Melbourne, Victoria}
\affiliation{Nagoya University, Nagoya}
\affiliation{Nara Women's University, Nara}
\affiliation{National Central University, Chung-li}
\affiliation{National United University, Miao Li}
\affiliation{Department of Physics, National Taiwan University, Taipei}
\affiliation{H. Niewodniczanski Institute of Nuclear Physics, Krakow}
\affiliation{Nippon Dental University, Niigata}
\affiliation{Niigata University, Niigata}
\affiliation{University of Nova Gorica, Nova Gorica}
\affiliation{Osaka City University, Osaka}
\affiliation{Osaka University, Osaka}
\affiliation{Panjab University, Chandigarh}
\affiliation{Peking University, Beijing}
\affiliation{University of Pittsburgh, Pittsburgh, Pennsylvania 15260}
\affiliation{Princeton University, Princeton, New Jersey 08544}
\affiliation{RIKEN BNL Research Center, Upton, New York 11973}
\affiliation{Saga University, Saga}
\affiliation{University of Science and Technology of China, Hefei}
\affiliation{Seoul National University, Seoul}
\affiliation{Shinshu University, Nagano}
\affiliation{Sungkyunkwan University, Suwon}
\affiliation{University of Sydney, Sydney NSW}
\affiliation{Tata Institute of Fundamental Research, Bombay}
\affiliation{Toho University, Funabashi}
\affiliation{Tohoku Gakuin University, Tagajo}
\affiliation{Tohoku University, Sendai}
\affiliation{Department of Physics, University of Tokyo, Tokyo}
\affiliation{Tokyo Institute of Technology, Tokyo}
\affiliation{Tokyo Metropolitan University, Tokyo}
\affiliation{Tokyo University of Agriculture and Technology, Tokyo}
\affiliation{Toyama National College of Maritime Technology, Toyama}
\affiliation{University of Tsukuba, Tsukuba}
\affiliation{Virginia Polytechnic Institute and State University, Blacksburg, Virginia 24061}
\affiliation{Yonsei University, Seoul}
  \author{K.~Abe}\affiliation{High Energy Accelerator Research Organization (KEK), Tsukuba} 
  \author{K.~Abe}\affiliation{Tohoku Gakuin University, Tagajo} 
  \author{I.~Adachi}\affiliation{High Energy Accelerator Research Organization (KEK), Tsukuba} 
  \author{H.~Aihara}\affiliation{Department of Physics, University of Tokyo, Tokyo} 
  \author{D.~Anipko}\affiliation{Budker Institute of Nuclear Physics, Novosibirsk} 
  \author{K.~Aoki}\affiliation{Nagoya University, Nagoya} 
  \author{T.~Arakawa}\affiliation{Niigata University, Niigata} 
  \author{K.~Arinstein}\affiliation{Budker Institute of Nuclear Physics, Novosibirsk} 
  \author{Y.~Asano}\affiliation{University of Tsukuba, Tsukuba} 
  \author{T.~Aso}\affiliation{Toyama National College of Maritime Technology, Toyama} 
% ...
\collaboration{The Belle Collaboration}

\noaffiliation

\begin{abstract}
We update our search for the lepton flavor violating 
$\tau^- \to \mu^- \gamma$ 
and $\tau^- \to e^- \gamma$ decays based on 535 fb$^{-1}$ 
of data accumulated at the Belle experiment. 
No signal is found and 
we set preliminary 90\% confidence level upper limits:
${\cal B}(\tau^- \to \mu^- \gamma) < 4.5\times 10^{-8}$ and
${\cal B}(\tau^- \to e^- \gamma) < 1.2\times 10^{-7}$.
\end{abstract}

\maketitle

\tighten

\section{Introduction}

To search for new physics beyond the Standard Model, we have been 
looking for the lepton flavor violating (LFV) $\TMG$ and $\TEG$ decays
in the Belle experiment~\cite{Belle} at the KEKB asymmetric-energy 
$e^+e^-$ collider~\cite{KEKB}.
Previously, we obtained the upper limits of 
${\cal B}(\TMG)<3.1\times10^{-7}$~\cite{TMG} and 
${\cal B}(\TEG)<3.9\times10^{-7}$~\cite{TEG} at the 90\% confidence
level (CL), 
using about 86 \fbi 
of data recorded at the $\Upsilon(4S)$ resonance. 
Later, the BaBar collaboration obtained the upper limits  
${\cal B}(\TMG)<6.8\times10^{-8}$~\cite{TMG.babar} and 
${\cal B}(\TEG)<1.1\times10^{-7}$~\cite{TEG.babar} with 232.2 \fbi of data.
Here we report our updated analysis with a data sample of 535 \fbi.

The Belle detector is a large-solid-angle magnetic
spectrometer that
consists of a silicon vertex detector (SVD),
a 50-layer central drift chamber (CDC), an array of
aerogel threshold \v{C}erenkov counters (ACC), 
a barrel-like arrangement of time-of-flight
scintillation counters (TOF), and an electromagnetic calorimeter
comprised of CsI(Tl) crystals (ECL) located inside 
a superconducting solenoid coil that provides a 1.5~T
magnetic field.  An iron flux-return located outside of
the coil is instrumented to detect $K_L^0$ mesons and to identify
muons (KLM).  The detector
is described in detail elsewhere~\cite{Belle}.

The basic analysis procedure is similar to our previous one \cite{TMG,TEG}.
The selection criteria for $\tau\rightarrow\mu\gamma/e\gamma$
are determined and optimized by 
examing Monte Carlo (MC) simulated singal and background (BG) events,
including the generic $\tau^+ \tau^-$,
$q\bar{q}$ $(q=u,d,s,c,b)$,
 Bhabha, $\mu^{+}\mu^{-}$, and two-photon events.
The BG $\tau^{+}\tau^{-}$ events are generated by the
KKMC/TAUOLA~\cite{KKMC} and
the response of the Belle detector is simulated by the GEANT3~\cite{GEANT3}
based program.

Photon candidates are defined as in Ref.~\cite{photonrecon}.
Muon candidates are identified by using a muon likelihood ratio,
${\cal L}_{\mu}$~\cite{muonID}, which is based on the difference between 
the range of the track
calculated from the particle momentum and that measured
by the KLM, which includes the value of $\chi^2$ formed from the KLM hit
locations with respect to the extrapolated track.
Identification of electrons is performed using
an electron likelihood ratio, 
${\cal L}_e$, based on the $dE/dx$ information
from the CDC, the ratio of the energy
deposited in the ECL to the
momentum measured by both the CDC and the SVD, 
the shower shape in the ECL, 
the hit information from the ACC
and the matching between the positions of the charged track
and the ECL cluster~\cite{electronID}.

\section{\boldmath $\tau\rightarrow\mu\gamma$}

\subsection{Event Selection}

We select events that include exactly two oppositely-charged tracks and 
at least one photon, consistent with $\tau^+\tau^-$ decays: 
one $\tau^{\pm}$ (signal side) decays to $\mu^{\pm} \gamma$ and the other 
(tag side) decays to a charged particle that is not a muon 
(denoted hereafter as $\notM$), neutrino(s) and any number of photons.

\begin{figure}[h]
\center{ \includegraphics[width=0.7\textwidth]{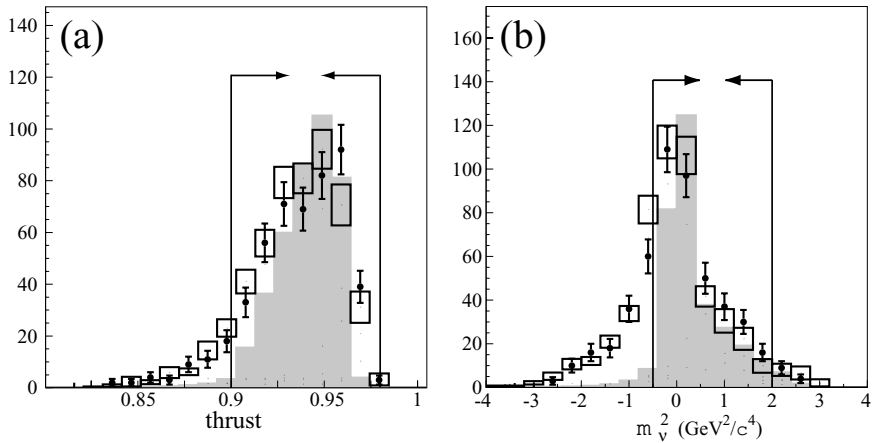} }
\caption{
(a) Length of the thrust vector and (b) $m_{\nu}^2$ distributions
for $\TMG$. Dots are data, open boxes show the BG Monte Carlo(MC) distribution
and shaded histograms are the signal MC.
Arrows indicate the selected region.
}
\label{cutmg}
\end{figure}

Each track should have a momentum $p^{\rm CM} <$ 4.5 GeV/$c$ 
in the center-of-mass (CM) frame and  
transverse component to the beam axis
$p_t >$ 0.1 GeV/$c$ within the detector fiducial region 
$ -0.866 < \cos\theta < 0.956$ to avoid  contamination
from Bhabha and $\mu^+\mu^-$ events.
(Hereafter all the variables defined in the CM frame have 
superscripts ``CM.'')
Each photon is required to have an energy $E_{\gamma}> 0.1$ GeV within 
the fiducial region. 
The total energy in the CM frame should be $E_{\rm total}^{\rm CM} < 10.5$ GeV 
to further suppress Bhabha and $\mu^{+}\mu^{-}$ events.
The magnitude of the thrust vector,
constructed from all selected charged tracks and photons above,
is required to be in the range from 0.90 to 0.98 
in order to suppress $\mu^{+}\mu^{-}$ 
and $q\bar{q}$ events.
(Fig.~\ref{cutmg}(a)).

For muon identification,
we require a likelihood ratio ${\cal L}_{\mu} >$ 0.95 and $p >$ 1.0~GeV/$c$. 
On the tag side, a track with ${\cal L}_{\mu} <$ 0.8 is defined as $\notmu$.
The photon that forms a $(\MG)$ candidate is required to have 
$E_{\gamma} >$ 0.5 GeV and $-0.602 < \cos\theta_{\gamma} < 0.829$ to 
remove spurious combinations of $\gamma$'s. 

The opening angle between the $\mu$ and $\gamma$ of $(\mu\gamma)$, 
0.4 $< \cos\theta^{\rm CM}_{\MG} <$ 0.8, is particularly powerful in rejecting 
$\tau^+\tau^-$ background, which contains $\pi^0$'s from $\tau$ decays.
The sum of the energies 
of the two charged tracks and the photon of the $(\MG)$, 
$E^{\rm CM}_{\rm sum}$, should be less than 9.0 GeV to reject 
$\mu^+\mu^-$ events. 
The opening angle between the two tracks should be greater than 90$^{\circ}$ 
in the CM frame while the opening angle between the $\mu$ 
and the boost direction of its mother
$\tau$ from the CM frame in the $\tau$ rest frame
is required to satisfy
$\cos\theta_{\mu\tau}<0.4$,
to remove combinations of $\mu$ and $\gamma$ from BG.

The following constraints on the momentum and the polar angle of the 
missing particle are imposed: 
$p_{\rm miss} >$ 0.4 GeV/$c$ and $-0.866 < \cos{\theta}_{\rm miss} <
0.956$. Here, $p_{\rm miss}$ is calculated by subtracting the momentum
of all charged tracks and photons from the beam momenta.
To remove the $\tau^+\tau^-$ background, a requirement on an opening
   between the tag-side track and the missing particle is applied,
0.4 $< \cos\theta^{\rm CM}_{{\rm miss}-\notM} <$ 0.98.
We calculate the missing mass squared on the tag side, $m_{\nu}^2=
(E_{\mu\gamma}-E_{\rm tag})^2-p_{\rm miss}^2$, 
where $E_{\mu\gamma}$ ($E_{\rm tag}$)
is the sum of the energy of the signal side $\mu$ and $\gamma$
(the sum of the energy of
all tag-side particles assuming $m_{\pi}$ for the tag-side track),
and then require $-1.0$ (GeV${}^2$/$c^4$)
$<m_{\nu}^2<2.0$ (GeV${}^2$/$c^4$), as shown in Fig.~\ref{cutmg}(b). 
Finally, a condition is imposed on the relation between $p_{\rm miss}$ and 
the mass squared of a missing particle, 
$m^2_{\rm miss}=E^2_{\rm miss} - p^2_{\rm miss}$: 
$p_{\rm miss} > -5 \times m^2_{\rm miss}-1$ and 
$p_{\rm miss} > 1.5 \times m^2_{\rm miss}-1$,
where $E_{\rm miss}$ is the sum of the beam energies minus the
sum of all visible energy and is calculated assuming the muon (pion)
mass for the charged track on the signal (tag) side,
$p_{\rm miss}$ is in GeV/$c$ and $m_{\rm miss}$ is in GeV/$c^2$.

\subsection{Background contribution}

After the selection requirements described in the previous subsection, 
the dominant BG source is $\tau^+\tau^-$ events
with the decay $\tau^\pm \to \mu^\pm \nu_{\mu} \nu_{\tau}$ and a
photon from initial
state radiation or beam BG.
Other sources are the radiative $\mu^+\mu^-$ process and 
$\tau^+\tau^-$ events with $\tau^\pm \to \pi^\pm \nu_{\tau}$.

Two variables are used to identify the signals: $\minv$, 
     the invariant mass of ($\mu\gamma$), and $\Delta E = E_{\mu\gamma} 
- E^{\rm CM}_{\rm beam}$, the energy difference between the $(\mu\gamma)$ 
     energy and the beam energy in the CM frame,
where the signal should have $\minv\sim$ $m_{\tau}$ and $\dE\sim0$. 
The resolutions in $\minv$ and $\dE$ are estimated by fitting 
 asymmetric Gaussians to 
the signal MC distributions giving
$\sigma^{\rm high/low}_{\minv}=14.49\pm0.10/24.24\pm0.13$ 
\mevpcs \ and $\sigma^{\rm high/low}_{\dE}=35.29\pm0.49/81.41\pm0.94$ \mev,
where $\sigma^{\rm high/low}$ means the standard deviation at the higher/lower
side of the peak. 

To compare the data and MC simulation, we examine a $\pm5\sigma$ region with
$1.65$ \gevpcs \ $<\minv<1.85$ \gevpcs \ and $-0.41$ \gev \
$<\dE<0.17$ \gev, as seen in 
Fig.~\ref{2d_mg}(a). 
A blind analysis method is taken: a slanting $\pm 3\sigma$ ellipse region, 
indicated by the dashed ellipse in Fig.~\ref{2d_mg}(a), 
whose detection efficiency 
is $\epsilon_{3\sigma}=6.05\%$, is covered till the final stage 
of the analysis. 

After the selections,
we find 71 events remaining in data and 
73.4$\pm 6.7$ events remaining in MC in the $\pm5\sigma$ region
 outside of the blinded ellipse. 
The latter is dominated by $\tau^{+}\tau^{-}$ events with initial state 
radiation, $\tau^+ \tau^- \gamma$,
and is comprised of $58.8\pm 4.3$ $(70.3\pm 4.7)$ 
$\tau^{+}\tau^{-}\gamma$ events, 
$13.1\pm 4.9$ $(15.0\pm 5.3)$  $\mu^{+}\mu^{-}\gamma$ events,
with incorrect $\mu$ identification,
and $1.6\pm 1.6$ $(3.2\pm 2.2)$  two-photon events,
where the numbers 
in the parentheses are the BGs that remain in the entire sample.

This background composition was understood 
in the previous analysis; 
the $\tau^{+}\tau^{-}\gamma$ process yields contributions in the 
$\Delta E<0$ region, while 
$\mu^{+}\mu^{-}\gamma$ events mostly have $\Delta E>0$. 
The latter rate is $\sim$20\% of the former, 
and one additional small contribution 
of $\sim$5\% is known to exist. 
This BG distribution is well represented by a combination of Landau and 
Gaussian functions, as found in Ref.~\cite{TEG}. 
The final event distribution in this data sample is found to follow well
the BG function; 
the $\mu^+\mu^-\gamma$ contribution is
$(20\pm 13) \%$
of the $\tau^+\tau^-\gamma$ contribution,
while the rest ($(5.2\pm 4.1) \%$) is from the two-photon 
process $e^+e^- \to e^+e^-\mu^+\mu^-$.

\subsection{Signal extraction}

\begin{figure}[h]
\center{ \includegraphics[width=0.4\textwidth]{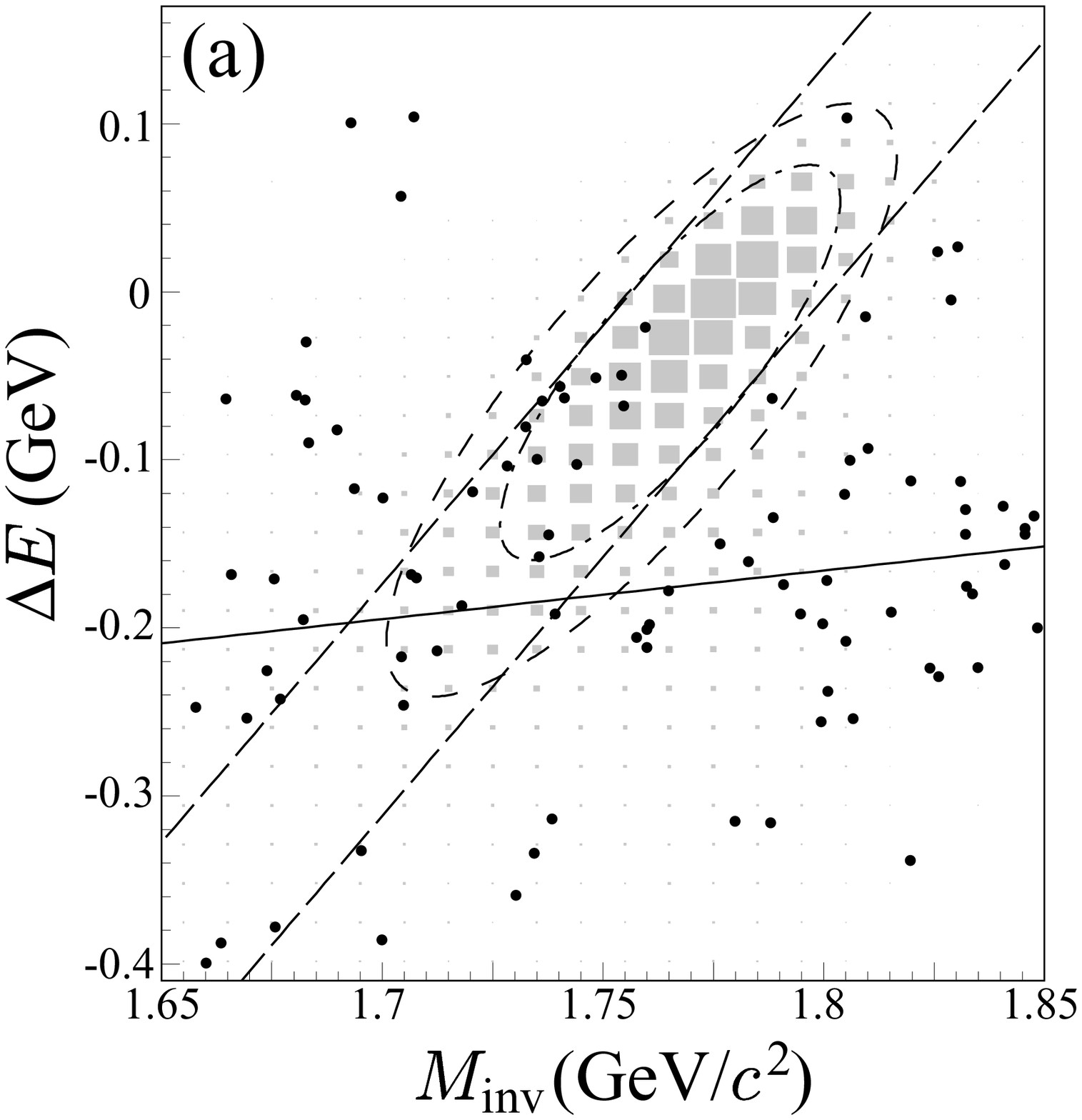}~~
         \includegraphics[width=0.4\textwidth]{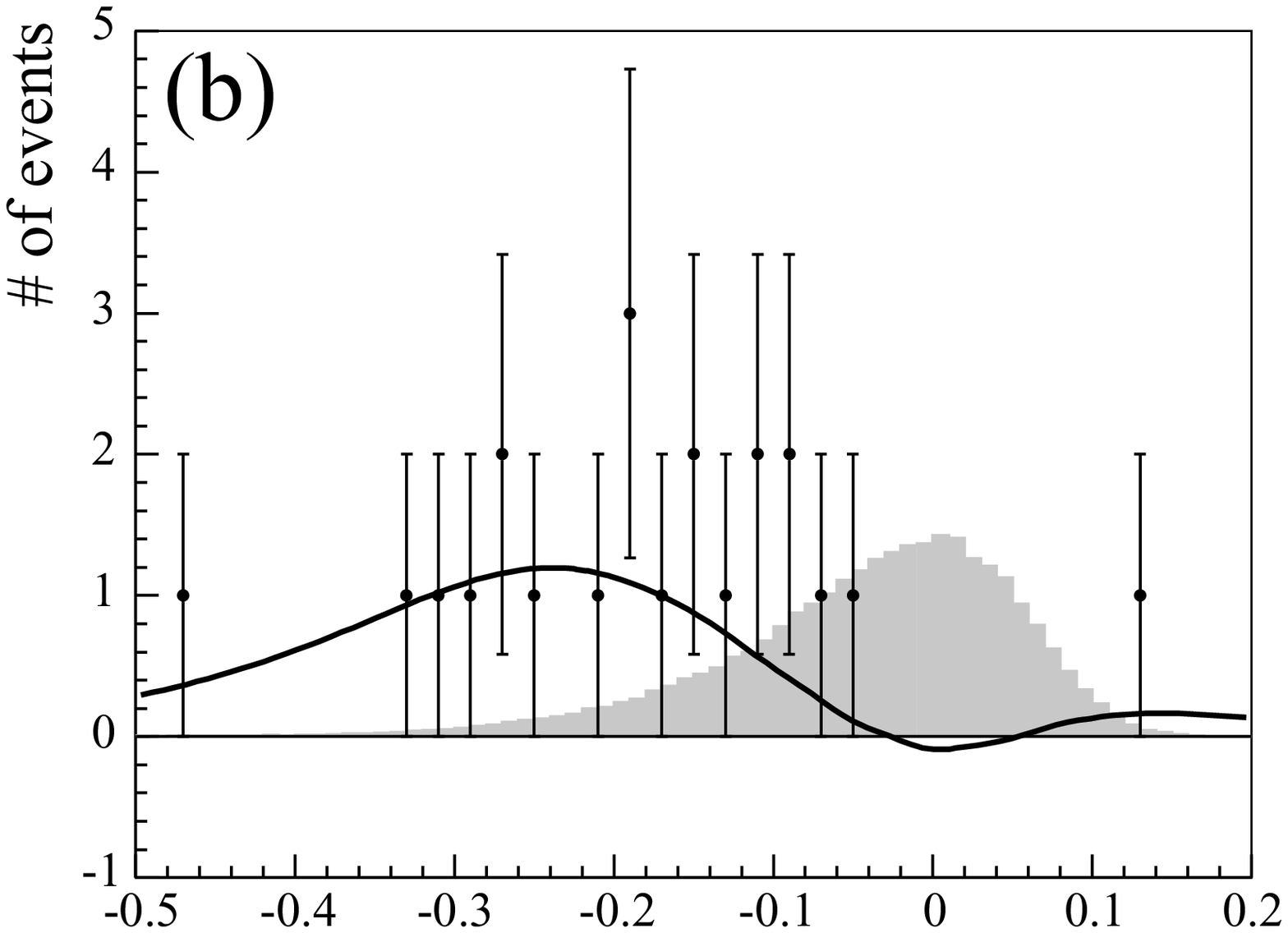} }
\caption{(a) $\minv$ -- $\dE$ distribution in the search for
$\TMG$ in the $\pm5\sigma$ region.
Dots are the data and shaded boxes indicate the signal MC.
The dashed ellipse shows the $3\sigma$ blinded region
and the dot-dashed ellipse is the $2\sigma$ signal region.
Dashed lines indicate the $\pm2\sigma$ band of the shorter ellipse axis,
projected onto the longer ellipse axis.
The solid line indicates the dense BG region.
(b) Data distribution within the $\pm2\sigma$ band.
Points with error bars 
are the data and the shaded histogram is the signal MC
assuming a branching ratio of $5\times10^{-7}$.
The solid curve shows the best fit.
}
\label{2d_mg}
\end{figure}

After unblinding, we find 23 and 94 data events in the blinded and 
$\pm5\sigma$ regions, respectively, while $15.0\pm3.1$
and $88.4\pm7.4$ events are expected from the MC. 
Figure~\ref{2d_mg}(a) shows the event distributions in the 
$\minv$--$\dE$ plane. 

In order to extract the number of signal events, 
we employ an unbinned extended maximum 
likelihood (UEML) fit  with the following likelihood function: 
\begin{equation}
  {\cal L}=\frac{e^{-(s+b)}}{N!}\prod_{i=1}^{N}
   \left(s S_i+b B_i \right).
\end{equation}
Here, $N$ is the number of observed events,
 $s$ and $b$ are the numbers of signal 
and BG events to be extracted, respectively, $S_i$ and $B_i$ are 
the signal and BG probability density functions (PDF), where $i$ indicates 
the $i$-th event, 
$S_i$ is obtained from the signal MC, and 
$B_i$ is the PDF for background mentioned above, whose distribution is
concentrated around $\Delta E\simeq -0.2$ GeV, as indicated by 
the solid line in Fig.~\ref{2d_mg}(a). 
To enhance the signal detection sensitivity 
and to avoid this dense BG region, 
we use a $\pm 2\sigma$ ellipse as the signal region for the UEML fit. 
The result of the fit is $s=-3.9$, $b=13.9$ with $N=10$.

Figure~\ref{2d_mg}(b) shows the event distribution within 
the $\pm 2\sigma$ band 
of the shorter ellipse axis, projected onto the longer ellipse axis, and 
the best fit curve. 
No events are found around the peak of the signal distribution. 
The negative $s$ value is consistent with no signal.

We examine the probability for obtaining this result and evaluate the 90\% 
CL upper limit using a toy MC simulation. 
The toy MC generates signal and BG events according to their PDFs 
fixing the expected number of BG events $(\tilde{b})$ at $\tilde{b}=b$ 
while varying the number of signal events ($\tilde{s}$). 
For every assumed $\tilde{s}$,  10,000 samples are generated following 
Poisson statistics with means $\tilde{s}$ and $\tilde{b}$ for the signal 
and BG, respectively; 
the signal yield $(s^{\rm MC})$ is evaluated by the UEML fit. 
To obtain the upper limit at 90\% CL ($\tilde{s}_{90}$) we take 
the $\tilde{s}$ value that gives a 90\% probability
 of $s^{\rm MC}$ to be larger 
than $\tilde{s}$.  
The probability to obtain $s\leq -3.9$ is 25\% in a case of a null 
true signal.
In other words, due to BG fluctuations
a negative $s$ value is possible with a large probability, 
although the physical signal rate is positive~\cite{OLDadd}.

The toy MC provides an upper limit on the signal at the 90\% CL as 
$\tilde{s}_{90}=2.0$ events from the result of the UEML fit. 
We then obtain the upper limit on the branching fraction 
${\cal B}_{90}(\tau\rightarrow\mu\gamma)$ 
at the 90\% CL as
\begin{equation}
 {\cal B}_{90}(\tau\rightarrow\mu\gamma)\equiv
\frac{\tilde{s}_{90}}{2\epsilon N_{\tau\tau}} < 4.1\times 10^{-8},
\end{equation}
where the number of $\tau$ pairs produced is $N_{\tau\tau} = 4.77 \times 10^8$,
and the detection efficiency for the $\pm 2\sigma$ ellipse region is 
$\epsilon= 5.07\%$.

The systematic uncertainties for the BG PDF shape increase
$\tilde{s}_{90}$ to 2.2.
The systematic uncertainties for $\epsilon$ arise from
the track reconstruction efficiency (2.0\%), 
the photon reconstruction efficiency (2.0\%), 
the selection criteria (2.2\%), 
the luminosity (1.4\%), 
the trigger efficiency (0.9\%), 
and the MC statistics (0.3\%). 
All errors are added in quadrature to yield the total uncertainty of 4.0\%. 
This uncertainty increases the upper limit of the branching ratio by
0.2\%~\cite{CLEO.mgeg}.
Since the angular distribution for $\tau\to\mu\gamma$ depends on 
the LFV interaction 
structure, we evaluate its effect on the result by assuming the maximum 
possible variation, $V\pm A$ interactions, rather than the 
uniform distribution so far assumed in the MC analysis. 
No appreciable effect is found for the upper limit.

Finally, the following upper limit on the branching ratio is obtained: 
\begin{equation}
 {\cal B}(\tau\rightarrow\mu\gamma)<4.5\times10^{-8} 
\hspace*{1 cm} {\rm at~the~90\% ~CL.}
\end{equation}
\vspace*{2 mm}

\section{\boldmath $\tau\rightarrow{e}\gamma$}

For $\tau\rightarrow e\gamma$ 
we use a procedure similar to that for 
$\tau\rightarrow\mu\gamma$. 

\subsection{Event Selection}

We examine a $\tau^+\tau^-$ sample, in which one $\tau$ goes to 
an electron and a photon, and the other $\tau$ decays to a charged particle, 
but not an electron ($\notE$), neutrino(s) and any number of photons.
The selection criteria are basically the same as those for $\tau\to\mu\gamma$, 
so below we describe only the differences.

An obvious difference is the replacement of the $\mu$ by an $e$ on the 
signal side, 
and using a $\notE$ veto rather than a $\notM$ veto on the tag side.
The electron on the signal side $(\EG)$ is required to have 
${\cal L}_{\rm e} >$ 0.90 and a momentum $p >$ 1.0 GeV/$c$, while the $\notE$ 
on the tag side should have ${\cal L}_{\rm e} <$ 0.1.
Minor differences in the kinematical selection include requirements
on the missing 
mass squared 
on the tag side and the opening angle between the tag-side track and the 
missing particle on the tag side: $-0.5$ (GeV${}^2$/$c^4$)  
$< m_{\nu}^2 < 2.0$ (GeV${}^2$/$c^4$), 
and 0.4 $< \cos\theta^{\rm CM}_{{\rm miss}-\notE} <$ 0.99. 
The other requirements are the same as those for $\tau\to\mu\gamma$.

The $M_{\rm inv}$ and $\Delta E$ resolutions are evaluated as 
$\sigma^{\rm high/low}_{\minv}=14.76\pm0.18/25.38\pm0.38$ \mevpcs \ and 
$\sigma^{\rm high/low}_{\dE}=35.66\pm0.62/89.98\pm1.72$ \mev, and then 
the $\pm5\sigma$ region over 
$1.65$ \gevpcs \ 
$<\minv<1.85$ \gevpcs \ and $-0.45$ \gev \ $<\dE<0.18$ \gev \ 
is assigned for the signal evaluation. 
A slanting $\pm3\sigma$ ellipse is also blinded. 

After the selection, we find 34 and $29.9\pm2.8$ events remaining 
in data and MC, respectively, in the $\pm5\sigma$ region, but excluding 
the blind one. 
The BG is from $\tau^{+}\tau^{-}\gamma$ events;
no other BG process is found 
to contribute.

\subsection{Signal extraction}

\begin{figure}[h]
\center{ \includegraphics[width=0.4\textwidth]{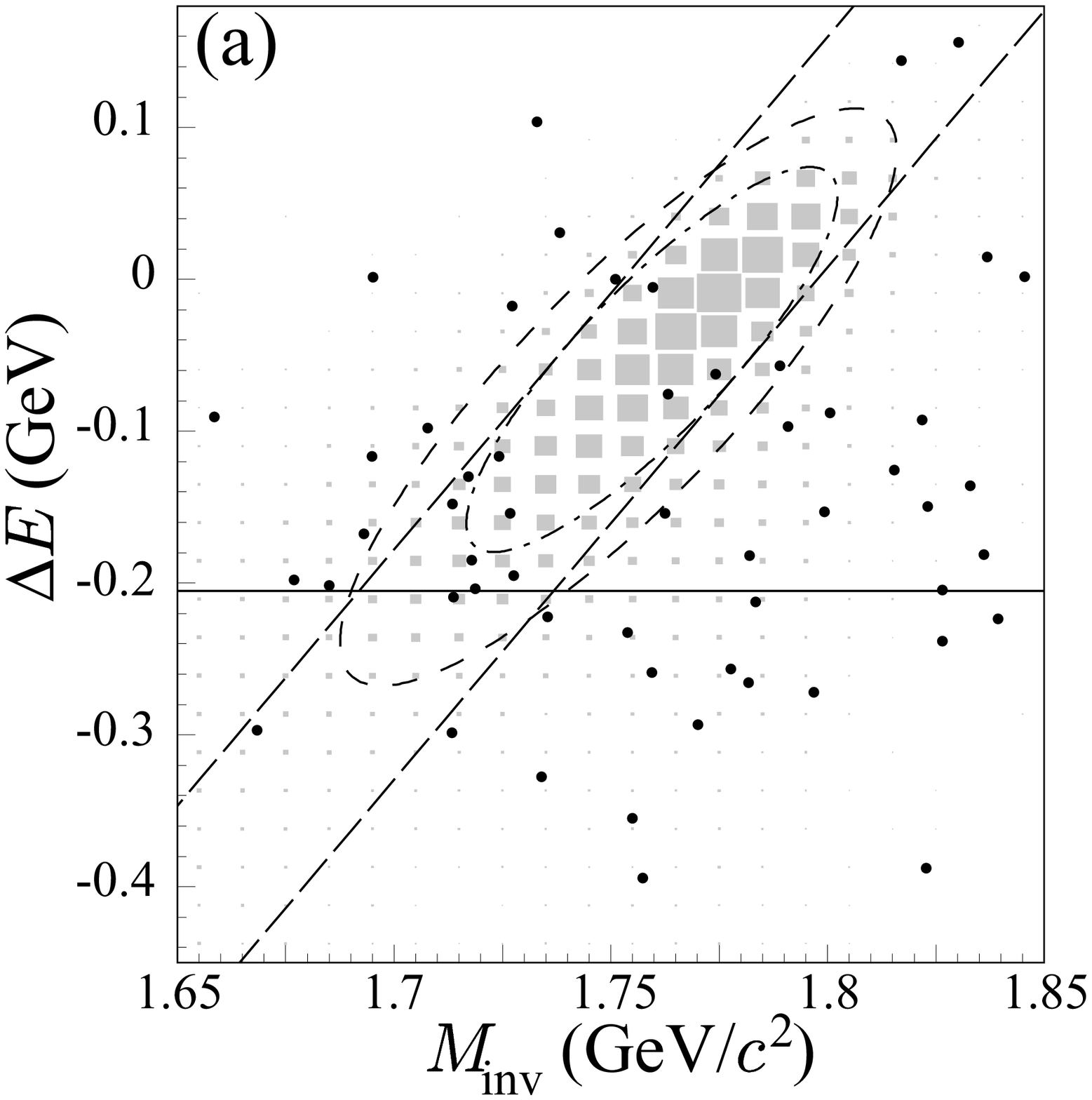}~~
         \includegraphics[width=0.4\textwidth]{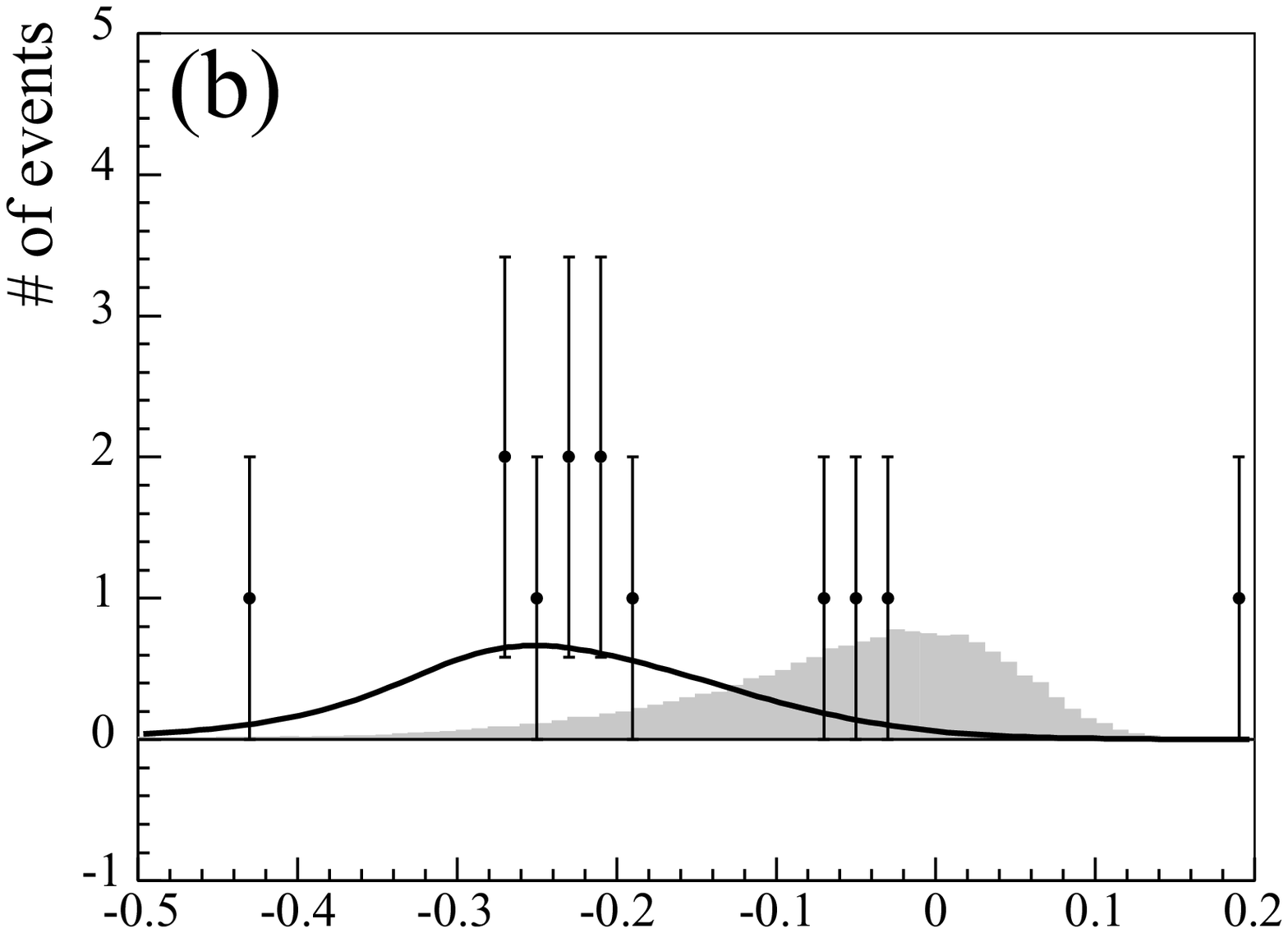} }
\caption{(a) $\minv$ -- $\dE$ distribution in the search for
$\TEG$ in the $\pm5\sigma$ region.
Dots are the data and shaded boxes indicate the signal MC.
The dashed ellipse shows the $3\sigma$ blinded region
and the dot-dashed ellipse is the $2\sigma$ signal region.
The dashed lines indicate the $\pm2\sigma$ band of the shorter ellipse axis,
projected onto the longer ellipse axis.
The solid line indicates the dense BG region.
(b) Data distribution within the $\pm2\sigma$ band.
Points with error bars are the data and 
the shaded histogram is the signal MC
assuming a branching ratio of $5\times10^{-7}$.
The solid curve shows the best fit.
}
\label{2d_eg}
\end{figure}

After opening the blind we find 13 and 55 data events in the blind and 
$\pm5\sigma$ region, respectively, while the MC predicts
$8.1\pm1.6$ and $42.8\pm 3.7$ events. 
Figure~\ref{2d_eg}(a) shows the event distribution 
in the $\minv$--$\dE$ plane. 

The signal extraction process is the same as that for $\tau\to\mu\gamma$, 
described in the former section. 
The BG is composed of  $(18 \pm 18)\%$ $e^+e^-\gamma$ (radiative
Bhabha), while the remainder is $\tau^+\tau^-\gamma$.
No other background source is found in MC. 
The UEML fit over the $\pm 2\sigma$ ellipse region results
 in $s=-0.14$, $b=5.14$ with $N=5$. 
The toy MC gives a probability of 48\% for obtaining $s\leq -0.14$ in 
the case of a null signal. 
Figure~\ref{2d_eg}(b) is the same as Fig.~\ref{2d_mg}(b), but for 
the $\tau\to e\gamma$ case. 
The upper limit of $\tilde{s}_{90}=3.3$ is obtained by the toy MC in 
the case of the UEML fit result.
The upper limit on the branching fraction is calculated as 
\begin{equation}
{\cal B}_{90}(\tau\rightarrow e\gamma)\equiv
\frac{\tilde{s}_{90}}{2\epsilon N_{\tau\tau}} < 11.7\times 10^{-8},
\end{equation}
where the detection efficiency for the $\pm 2\sigma$ ellipse region is 
$\epsilon= 2.99\%$. 

The systematic uncertainties are essentially the same as those for 
$\tau\to\mu\gamma$: minor differences are the selection criteria (2.5\%), 
and the trigger efficiency (2.0\%). 
The total uncertainty amounts to 4.5\%, and it increases the upper limit 
of the branching ratio by 0.2\%. In addition,
the systematic uncertainties for the BG PDF shape increase
$\tilde{s}_{90}$ to 3.4.
Taking into account this systematic error, we obtain 
the 90\% CL upper limit,
\begin{equation}
{\cal B}(\tau\rightarrow{e}\gamma) < 12.0\times10^{-8}.
\end{equation}

\section{Summary}

We updated our LFV searches for $\TMG$ and $\TEG$ decays, based on 
535 fb$^{-1}$ of data, i.e. with about six times higher statistics than 
before: 
The resulting upper limits on the branching fractions are 
\begin{eqnarray}
 {\cal B}(\TMG) &<& 4.5 \times 10^{-8},\\
 {\cal B}(\TEG) &<& 12.0 \times 10^{-8}
\end{eqnarray}
at the 90\% CL. \\ 

For the $\tau\rightarrow\mu\gamma$ search,
we obtain $\epsilon=5.07$\% and $N_{\rm obs}=10$
in the region where the upper limit is evaluated,
while we had
 $\epsilon=12.0$\% and $N_{\rm obs}=54$
in our previous analysis with the 86 fb${}^{-1}$ data sample.
Similarly, for the $\tau\rightarrow{e}\gamma$ search,
we obtain $\epsilon=2.99$\% and $N_{\rm obs}=5$
compared to
 $\epsilon=6.39$\% and $N_{\rm obs}=20$ previously.
Both modes have  about 6 times better sensitivity
than the previous analyses.
We have also estimated the improvement in our sensitivity
for the old data sample.
The result would be only 2.5 times better 
if we applied the methods used in our previous analysis
to the current data sample.

\section{Acknowledgments}

We thank the KEKB group for the excellent operation of the
accelerator, the KEK cryogenics group for the efficient
operation of the solenoid, and the KEK computer group and
the National Institute of Informatics for valuable computing
and Super-SINET network support. We acknowledge support from
the Ministry of Education, Culture, Sports, Science, and
Technology of Japan and the Japan Society for the Promotion
of Science; the Australian Research Council and the
Australian Department of Education, Science and Training;
the National Science Foundation of China and the Knowledge 
Innovation Program of the Chinese Academy of Sciencies under 
contract No.~10575109 and IHEP-U-503; the Department of Science 
and Technology of India; the BK21 program of the Ministry of Education of
Korea, and the CHEP SRC program and Basic Research program 
(grant No. R01-2005-000-10089-0) of the Korea Science and
Engineering Foundation; the Polish State Committee for
Scientific Research under contract No.~2P03B 01324; the
Ministry of Science and Technology of the Russian
Federation; the Slovenian Research Agency;  
the Swiss National Science Foundation; the National Science Council and
the Ministry of Education of Taiwan; and the U.S.\
Department of Energy.


\begin{thebibliography}{99}
\bibitem{OLDadd}
We restricted $s$ to be positive in the previous analysis. 
Due to BG fluctuations, however, $s$ can be negative as we found in this 
case. Our previous upper limit is then reduced by a factor of $\simeq 3$ 
when negative $s$ is allowed: the revised upper limit changes from
$Br(\tau\to\mu\gamma)< 3.1\times 10^{-7}$ to $0.9\times 10^{-7}$.
\end{thebibliography}
\end{document}